\documentclass[doublecol]{epl2}

\usepackage{graphicx}
\usepackage{amsmath,amsfonts,amssymb}

\newcommand{\mean}[1]{\left \langle #1 \right \rangle}
\newcommand{\dd}{\,\mathrm{d}}
\newcommand{\ym}{y_\text{m}}
\newcommand{\xm}{x_\text{m}}
\newcommand{\N}{{\cal N}}
\newcommand{\li}{\lambda_\text{i}}
\newcommand{\lf}{\lambda_\text{f}}
\newcommand{\ki}{k_\text{i}}
\newcommand{\kf}{k_\text{f}}
\newcommand{\kb}{k_\text{B}}
\newcommand{\bi}{b_\text{i}}
\newcommand{\yi}{y_\text{i}}
\newcommand{\sm}{s_\text{m}}
\newcommand{\st}{s_\text{tot}}
\newcommand{\ssys}{s_\text{sys}}
\newcommand{\ssm}{S_\text{sys}}
\newcommand{\stm}{S_\text{tot}}
\newcommand{\logn}[1]{\ln\left( #1 \right)}

\title{Extracting work from a single heat bath through feedback}

\author{D. Abreu \and U. Seifert}
\shortauthor{D. Abreu and U. Seifert}

\institute{II. Institut f\"ur Theoretische Physik, Universit\"at Stuttgart - 70550 Stuttgart, Germany}


\abstract{Work can be extracted from a single heat bath if additional information is available. For the paradigmatic case of a Brownian particle in a harmonic potential, whose position has been measured with finite precision, we determine the optimal protocol for manipulating the center and stiffness of the potential in order to maximize this work in a finite-time process. The bound on this work imposed by a generalized second law inequality involving information can be reached only if both position and stiffness of the potential are controlled and the process is quasistatic. Estimates on the power delivered by such an ``information machine'' operating cyclically follow from our analytical results.}

\begin{document}

\maketitle

\section{Introduction}
The idea of extracting work from a single heat bath has been investigated for a long time as a putative challenge to the validity of the second law of thermodynamics. Thought experiments such as Maxwell's demon or Szilard's engine, which supposedly violate the second law, were resolved by taking into account the thermodynamic processing of the acquired information through Landauer's principle, which quantifies the cost of information erasure\cite{leff03,maru09}.
Whether it is the acquisition or the deletion of this information that compensates for the gained work still generates debate \cite{saga09,dill09}.
The fact, however, that measurements can help to extract work from a single thermal bath has been widely investigated, mainly in theory for the classical \cite{alla08,cao09,suzu09,mara10} as well as for the quantum regime \cite{alla00,scul01,alla04,saga08,jaco09,camp10}, and validated by a recent experiment \cite{toya10a}.
This idea is based on the strong link between statistical physics and information theory \cite{jayn57,schl80,touc04,broe10}, in particular how the information-theoretic concepts of relative entropy or mutual information can be related to the thermodynamic notion of work \cite{kawa07,gome08,pele10}. The generalization of the Jarzynski relation \cite{jarz97} formulated in \cite{saga10} and the subsequent refinements of the fluctuation theorems \cite{horo10,ponm10} take information explicitly into account as a way to extract more work than the corresponding free energy difference from a non-equilibrium process.

In this Letter, we address the question of how to extract within a finite-time the maximum amount of work exploiting the information gained from one measurement, using the framework of stochastic thermodynamics \cite{seif07,seki10}. As a model system, we consider a Brownian particle in a harmonic potential as in an optical trap. Several groups have used such colloidal particles in optical tweezers for experiments illustrating and testing the concepts of stochastic thermodynamics \cite{wang02,carb04,trep04,spec07,blic06,gome10}. We determine the optimal protocol \cite{schm07,schm09a,then08,espo10a,geig10} for two cases where we either control only the center of a trap or additionally its stiffness. Only the latter case allows a complete conversion of all the information gained by a position measurement into work. Finally, we estimate the power of cyclic engines based on such optimal protocols.

\section{Dynamics}
We study the dynamics of a one dimensional Brownian particle in a single heat bath of temperature $T$, trapped in a harmonic potential 
  \begin{equation}
  V(x,\tau) = k(\tau)(x-\lambda(\tau))^2/2
  \label{pot}
  \end{equation}
and driven into non-equilibrium by controlling the time-dependence of the position $\lambda(\tau)$ and the stiffness $k(\tau)$ of this potential, from the initial values $\li\equiv0$ and $\ki$ to the end values $\lf$ and $\kf$ during the total time $t$. We call \textit{system} the particle in the harmonic potential and \textit{medium} the surrounding heat bath. We neglect inertial effects such that the dynamics is governed by the overdamped Langevin equation
  \begin{equation}
  \dot{x} = -\mu\partial_xV(x,\tau)  + \zeta(\tau),
  \label{langevin}
  \end{equation}
the dot representing the time derivative and $\mu$ the mobility of the particle. The thermal white noise $\zeta(\tau)$ with
  \begin{align}
  \mean{\zeta(\tau)} &= 0, \\
  \mean{\zeta(\tau)\zeta(\tau')} &= 2\mu \kb T\delta(\tau-\tau'),
  \end{align} 
obeys the usual fluctuation-dissipation relation, the brackets $\mean{...}$ representing the average of a quantity.

Throughout the paper, we set four quantities to unity: the temperature $T$, the Boltzmann constant $\kb$, the mobility $\mu$, and the initial potential stiffness $\ki$. This corresponds to a scaling of energies by a factor $\kb T$, entropies by $\kb$, times by $1/\mu\ki$, and lengths by $\sqrt{\kb T/\ki}$. This rescaling does not reduce the generality of our results, given that these four quantities are independent units.

Since we want to analyze mean values of the work, we will not use the Langevin equation \eqref{langevin} but rather the corresponding Fokker-Planck equation, which with our units reads
  \begin{align}
  \partial_{\tau} p(x,\tau)&=\partial_x\left[\partial_xV(x,\tau)\,p(x,\tau)+\partial_x p(x,\tau)\right] \nonumber \\
  &\equiv -\partial_x j(x,\tau),
  \label{fokkerplanck}
  \end{align}
where $p(x,\tau)$ is the probability distribution function (PDF) for the position of the particle and $j(x,\tau)$ the probability current. We further make the ansatz of a Gaussian PDF
  \begin{align}
  p(x,\tau)& = \frac{1}{\sqrt{2\pi y^2(\tau)}}\exp\left(-\frac{(x-b(\tau))^2}{2y^2(\tau)}\right)  \nonumber \\
  &\equiv \N_x(b(\tau),y^2(\tau))
  \label{gauss}
  \end{align}
with average $b(\tau)$ and variance $y^2(\tau)$ defining the compact notation $ \N_x(b,y^2)$.
By inserting eqs. \eqref{pot} and \eqref{gauss} into the Fokker-Planck equation \eqref{fokkerplanck} and identifying the coefficients of the resulting polynomial in $x$, we obtain the two equations of motion,
  \begin{align}
  k(\tau)&=\frac{1}{y^2(\tau)}-\frac{\dot{y}(\tau)}{y(\tau)}, \label{fpk}\\
  \lambda(\tau)&=b(\tau) +\frac{\dot{b}(\tau)}{k(\tau)}, \label{fpl}
  \end{align}
which relate the evolution of the control parameters $\lambda$ and $k$ to those of the two moments of the distribution $b$ and $y$. Eqs. \eqref{fpk} and \eqref{fpl} allow us to transform the problem of finding the optimal protocols $\lambda^*(\tau)$ and $k^*(\tau)$ into the one of determining the optimal functions $b^*(\tau)$ and $y^*(\tau)$, which turns out to be much simpler.

\section{Thermodynamics}
On a single trajectory $x(\tau)$, the first law holds as \cite{seki10}
  \begin{equation}
  w[x(\tau)] = \Delta V[x(\tau)]+ q[x(\tau)],
  \label{1hstr}
  \end{equation}
where $w>0$ represents work applied in the system, which will either lead to an increase $\Delta V\equiv V(x(t),t)-V(x(0),0)$ of the energy of the particle or be transmitted to the surrounding medium as heat $q$. In the following, we will talk of \textit{extracted work} for $w<0$. The maximum extracted work thus corresponds to the minimum applied work. Following \cite{seif05a}, we relate the heat $q[x(\tau)]$ to the entropy change $\Delta \sm$ of the medium 
  \begin{equation}
  q[x(\tau)] \equiv \Delta \sm[x(\tau)] \equiv \Delta\st[x(\tau)] - \Delta\ssys[x(\tau)]
  \label{qsm}
  \end{equation}
where $\Delta\st$ is the total entropy change and $\Delta\ssys$ the change in the entropy of the system.
By putting eq. \eqref{qsm} into the first law \eqref{1hstr}, we obtain
  \begin{equation}
  w[x(\tau)] = \Delta V[x(\tau)] + \Delta\st[x(\tau)] - \Delta\ssys[x(\tau)],
  \label{worksingle}
  \end{equation}
which expresses the work $w$ applied along a trajectory $x(\tau)$ as a function of the change of three quantities. We will minimize the mean total work
  \begin{equation}
  W \equiv \mean{w[x(\tau)]} = \int w[x(\tau)]p[x(\tau)]\dd [x(\tau)],
  \label{meanwork}
  \end{equation}
where the integration is over all possible trajectories $x(\tau)$ with the associated probability $p[x(\tau)]$.
We will not directly compute \eqref{meanwork} but instead use the fact that we can easily average the three terms on the right-hand side of eq. \eqref{worksingle} for time-dependent Gaussian distributions \eqref{gauss}.

The mean energy change $\Delta E\equiv\mean{V(x,t)}-\mean{V(x,0)}$ of the particle after the process is given by
  \begin{align}
  \Delta E &= \int p(x,t) V(x,t) \dd x - \int p(x,0) V(x,0) \dd x  \nonumber\\
  &= \frac{\kf}{2}\left[(b(t)-\lf)^2+y^2(t)\right]-\frac{1}{2}[b^2(0)+y^2(0)]
  \label{deltae}
  \end{align}
with the initial conditions $\li=0$ and $\ki=1$. The mean entropy of the system is 
the usual time-dependent Gibbs entropy $\ssm (\tau) \equiv \mean{\ssys[x(\tau)]} = -\int p(x,\tau)\ln p(x,\tau)\dd x $. The mean total entropy change of the system thus is
  \begin{equation}
  \Delta \ssm = \logn{y(t)/y(0)}.
  \label{deltasm}
  \end{equation}
Once again following \cite{seif05a}, we find that
  \begin{equation}
  \dot{S}_\text{tot} \equiv \mean{\dot{s}_\text{tot}} = \int  \frac{j^2(x,\tau)}{p(x,\tau)} \dd x = \dot{b}^2+\dot{y}^2
  \end{equation}
so that, by integrating this rate over the duration $t$ of the process, we obtain the mean change of the total entropy
  \begin{equation}
  \Delta\stm = \int_0^t(\dot{b}^2+\dot{y}^2)\dd\tau.
  \label{deltastot}
  \end{equation}

The mean total work follows by averaging eq. \eqref{worksingle} using eqs. \eqref{deltae}, \eqref{deltasm}, and \eqref{deltastot} as
  \begin{align}
  W =&\frac{\kf}{2}\left[(b(t)-\lambda_\mathrm{f})^2 + y^2(t)\right]-\frac{1}{2}\left[b^2(0) + y^2(0)\right]\nonumber \\
  &-\logn{\frac{y(t)}{y(0)}}+\int_0^t (\dot{b}^2 + \dot{y}^2) \dd \tau.
  \label{work}
  \end{align}
In order to optimize expression \eqref{work}, we have to identify the initial distribution $\N_x(b(0),y^2(0))$. If the initial state is the thermal one, we have $b(0)=0$ and $y(0)=1$. However, if we first perform a measurement, the initial distribution will depend on its outcome and allow us to drive our system such that work can be extracted. We will show that the second law still holds in the generalized form \cite{kawa07,saga08}
  \begin{equation}
   W \geq \Delta F -I,
   \label{2ndlaw}
  \end{equation}
where
  \begin{equation}
  F \equiv -\logn{\int \exp\left(-V(x)\right)\dd x} = -\frac{1}{2}\logn{\frac{2\pi}{k}}
  \label{free}
  \end{equation}
is the free energy of the system and $I$ is the information acquired through the measurement as defined in the next section.

\section{Measurement}
The state of the system before the time $\tau=0$ of the measurement is represented by the PDF $p_\text{i}(x)\equiv\N_x(\bi,\yi^2)$. We assume that the measurement is instantaneous and that its outcome $\xm$ is distributed around the true position $x$ as
  \begin{equation}
  p(\xm |x) = \N_{\xm }(x,\ym^2),
  \label{pxmx}
  \end{equation}
where $\ym$ is the precision of the measurement. From eq. \eqref{pxmx} we can extract the PDF of the measurement outcome
  \begin{equation}
  p(\xm) = \int p_\text{i}(x)p(\xm|x) \dd x = \N_{\xm}(\bi,\yi^2+\ym^2).
  \label{pxm}
  \end{equation}
Through Bayes' theorem, $p(x|\xm)p(\xm) = p_\text{i}(x)p(\xm|x)$, we obtain the conditional PDF for the true position $x$ of the particle for a measured $\xm$ as
  \begin{equation}
  p(x|\xm) = \N_{x}\left(\frac{\xm \yi^2+\bi\ym^2}{\yi^2+\ym^2},\frac{\yi^2\ym^2}{\yi^2+\ym^2}\right).
  \label{pxxm}
  \end{equation}
The distribution $p(x|\xm)$ is the initial distribution $\N_x(b(0),y^2(0))$ that we have to use in eq. \eqref{work} if we have performed a measurement.

A quantity of interest in our problem is the Kullback-Leibler distance or relative entropy \cite{cove06} 
  \begin{align}
  &I(\xm) \equiv \int p(x|\xm)\logn{\frac{p(x|\xm)}{p(x)}}\dd x   \nonumber\\
  &= \frac{1}{2}\logn{1+\frac{\yi^2}{\ym^2}}+\frac{\yi^2}{2(\yi^2+\ym^2)}\left(\frac{(\xm-\bi)^2}{\yi^2+\ym^2}-1\right) \label{kldist}
  \end{align}
between the distribution $p(x|\xm)$ and $p(x)$. For a fixed measurement outcome $\xm$, this distance quantifies the distinguishability of the two distributions. If we average eq. \eqref{kldist} over all possible measurement results $\xm$ distributed with \eqref{pxm}, we obtain the mutual information
  \begin{equation}
  \bar{I} \equiv \int p(\xm)I(\xm) \dd\xm = \frac{1}{2}\logn{1+\frac{\yi^2}{\ym^2}}, \label{info}
  \end{equation}
where the bar denotes the average over $\xm$. The information that appears in eq. \eqref{2ndlaw} is given by eq. \eqref{kldist} for a fixed measurement outcome $\xm$ and by eq. \eqref{info} if we average over $\xm$, respectively.

\section{Instantaneous processes}
We first illustrate the generalized second law \eqref{2ndlaw} and show that by measuring once the position of the particle, we can extract work by moving the trap even though there is no free energy difference involved. We instantaneously move the potential from $\li=0$ to the final position $\lf(\xm)$ according to the measurement outcome while keeping its stiffness constant ($\kf=\ki=1$). In particular we want to find the optimal position $\lf^*(\xm)$ which minimizes the applied work, \textit{i.e.}, maximizes the extracted work.

\begin{figure}[t]
  \centering
  \includegraphics[width=\linewidth]{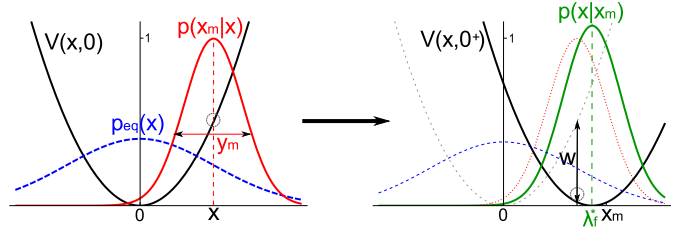}
  \caption{Optimal instantaneous protocol. Left: Initially, the potential is centered at $\li =0$ (black). The particle is in thermal equilibrium with its position $x$ distributed according to $p_\text{eq}(x)$ (blue, dashed). The measurement outcome $\xm$ is distributed with $p(\xm|x)$ (red) with precision $\ym$ around the real position $x$. Right: The optimal protocol consists in instantaneously displacing the potential to the position $\lf^*$, which is the center of the distribution $p(x|\xm)$ (green) and yields the work $w$, in average equal to \eqref{workinst}.}
  \label{Instant}
  \end{figure}

For a system initially in thermal equilibrium, we have $\bi=0$ and $\yi=1$ in the PDFs \eqref{pxm} and \eqref{pxxm}.
The work applied to the system is just the energy difference between final and initial state
  \begin{equation}
  w(x,\xm) = \frac{(x-\lf(\xm))^2-x^2}{2}.
  \end{equation}
For a fixed measurement outcome $\xm$, the mean work is given by 
  \begin{align}
  W(\xm) &\equiv \int p(x|\xm)w(x,\xm) \dd x \nonumber\\
  &= \frac{1}{2}\left[\left(\frac{\xm}{1+\ym^2}-\lf(\xm)\right)^2-\frac{\xm^2}{(1+\ym^2)^2}\right],
  \label{workinho}
  \end{align}
which becomes minimal for
  \begin{equation}
  \lf^*(\xm)=\frac{\xm}{1+\ym^2}.
  \label{optl}
  \end{equation}
This optimal value corresponds to the center of the distribution $p(x|\xm)$, as shown in fig. \ref{Instant}. 
The optimal work $W^*(\xm)$ is obtained by putting eq. \eqref{optl} into eq. \eqref{workinho} and is always negative, but still obeys the inequality \eqref{2ndlaw} since we have
  \begin{equation}
  W^*(\xm) + I(\xm) = \frac{-1}{2(1+\ym^2)}+\frac{1}{2}\logn{1+\frac{1}{\ym^2}} > 0 = \Delta F.
  \end{equation}
This inequality holds for fixed $\xm$ and then also averaged over $\xm$. Equality would be reached for $\ym\to\infty$, \textit{i.e.}, for an extremely imprecise measurement, for which we do not gain any work.
If we average $W^*(\xm)$ over all measurement outcomes $\xm$ distributed with \eqref{pxm}, we obtain the mean optimal total work
  \begin{equation}
  \overline{W^*} \equiv \int p(\xm)W^*(\xm)\dd\xm= -\frac{1}{2(1+\ym^2)}.
  \label{workinst}
  \end{equation}
For an error-free measurement, we would have $\ym=0$ and consequently $\overline{W^*}=-1/2$, which means that we were able to extract the total mean energy of the particle by applying an instantaneous shift of the potential following this ideal measurement. However, the mean amount of information \eqref{info} acquired through the measurement is always strictly greater than the mean extracted work \eqref{workinst}. We will be able to convert more information into work only if we allocate a finite-time to the process.

\section{Optimal finite-time protocols}
We now investigate optimal processes occurring in a finite-time $t$ and for which the end values $\lf$ and $\kf$ of the control parameters are assumed to be fixed. The setup is similar to the one in \cite{schm07} with the main difference that we first perform a measurement before driving the potential, which allows us to extract work from the system. 

In the following, the system is in thermal equilibrium before the measurement so that we have $\bi=0$ and $\yi=1$ and consequently 
\begin{equation}
 b(0)=\frac{\xm}{1+\ym^2} ~~\text{and}~~ y^2(0)=\frac{\ym^2}{1+\ym^2} \label{bzero}
\end{equation}
for the initial distribution \eqref{pxxm} after a measurement. We determine the optimal functions $b^*(\tau)$ and $y^*(\tau)$ and through eqs. \eqref{fpk} and \eqref{fpl} the optimal protocols $\lambda^*(\tau)$ and $k^*(\tau)$ that minimize the work \eqref{work} in two different cases.

\subsection{Case 1: Constant stiffness}
We first analyze the case in which the stiffness $k(\tau) = k_\text{i} = 1$ is constant. The subscript $_{1}$ will denote results obtained under this assumption. In this case, eq. \eqref{fpk} is simplified and we have
  \begin{equation}
  \int_0^t \dot{y}^2\dd\tau = \logn{\frac{y(t)}{y(0)}}-\frac{y^2(t)-y^2(0)}{2}
  \end{equation}
so that all terms depending on $y$ in eq. \eqref{work} cancel. The total work then assumes the form
  \begin{equation}
  W_{1} = \frac{(b(t)-\lambda_\mathrm{f})^2-b^2(0)}{2}+\int_0^t\dot{b}^2\dd\tau
  \label{workao}
  \end{equation}
that consists of one boundary term and one integral term with $b(0)$ given by eq. \eqref{bzero}. By optimizing first the integral term with the Euler-Lagrange equations, we find that $b(\tau)$ must be a linear function. We therefore optimize eq. \eqref{workao} with respect to the final value $b(t)$ and find the expression
  \begin{equation}
  b^*_{1}(t) = \frac{2b(0)+\lf t}{2+t},
  \end{equation}
from which we can determine the optimal protocol $\lambda_1^*(\tau)$ through eq. \eqref{fpl} and the optimal work $W^*_{1}(\xm)$ by inserting it into eq. \eqref{workao}. If we add this optimal work to the information \eqref{kldist}, we obtain
\begin{equation}
W^*_{1}(\xm)+I(\xm)= -\ln y(0) + \frac{y^2(0)-1}{2}+\frac{(b(0)-\lf)^2}{2+t}
\end{equation}
which is strictly positive due to $0\leq y(0)<1$. The second law \eqref{2ndlaw} is therefore respected since $\Delta F=0$. In every case though, the extracted work is strictly smaller than the gained information \eqref{kldist}, even in the quasistatic limit. We will see that we can extract work corresponding to the full information if we also control the stiffness of the potential. 

Using eq. \eqref{pxm} to average over all possible measurement results $\xm$, we find the mean total optimal work
  \begin{align}
  \overline{W^*_1} &\equiv \int p(\xm) W_1^*(\xm)\dd\xm \nonumber \\
 &=\frac{\lambda_\mathrm{f}^2}{2+t}-\frac{t}{2(2+t)}\frac{1}{1+y_\mathrm{m}^2}
  \label{wokstmit}
  \end{align}
consisting of two terms, where the first one corresponds to the one obtained without measurement in \cite{schm07}. The second one is always negative and corresponds to the work that can be gained through one measurement at constant stiffness. For $t=0$, we have $\overline{W^*_1}(t=0)=\lf^2/2$: there is only one way to go from $\li=0$ to a fixed $\lf$, namely an instantaneous jump, and we do not have time to take advantage of the acquired information - in this case, the measurement is  useless. In the quasistatic limit, $t\to\infty$, the first term vanishes and we can extract the work
  \begin{equation}
  \overline{W_1^*}(t\to\infty) = -\frac{1}{2(1+\ym^2)},
  \label{workkonst}
  \end{equation}
which is the same expression as the one found for instantaneous processes in eq. \eqref{workinst}, as there is no dissipation in both cases. However, the difference here is that we do not need to adapt the end position of the potential $\lf$ to the measurement outcome as it was the case in order to derive eq. \eqref{workinst}. We can rather extract work for any fixed value of $\lf$ provided we allocate enough time $t>2(1+\ym^2)\lf^2$ to the process. 

For the particular value $\lf=0$, \textit{i.e.}, imposing a final return to the initial potential, eq. \eqref{wokstmit} shows that work can be extracted in average for any total time $t$ of the protocol. In this case, the optimal work $W^*_{1}$ is equal to
\begin{equation}
 W^*_{1} = -\frac{t}{2+t}\frac{b^2(0)}{2},
\end{equation}
which means that we extract work from any initial state and any outcome of the measurement.

\subsection{Case 2: Control of both parameters}
\begin{figure}[t]
  \centering
  \includegraphics[width=0.85\linewidth]{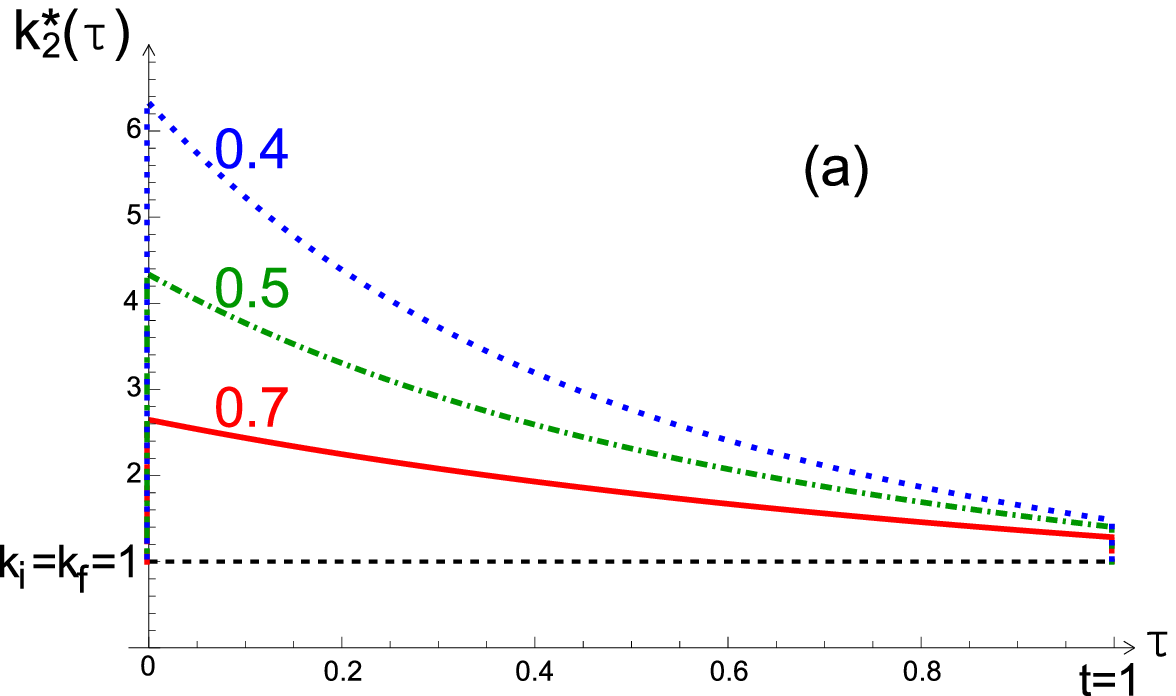}
  \includegraphics[width=0.85\linewidth]{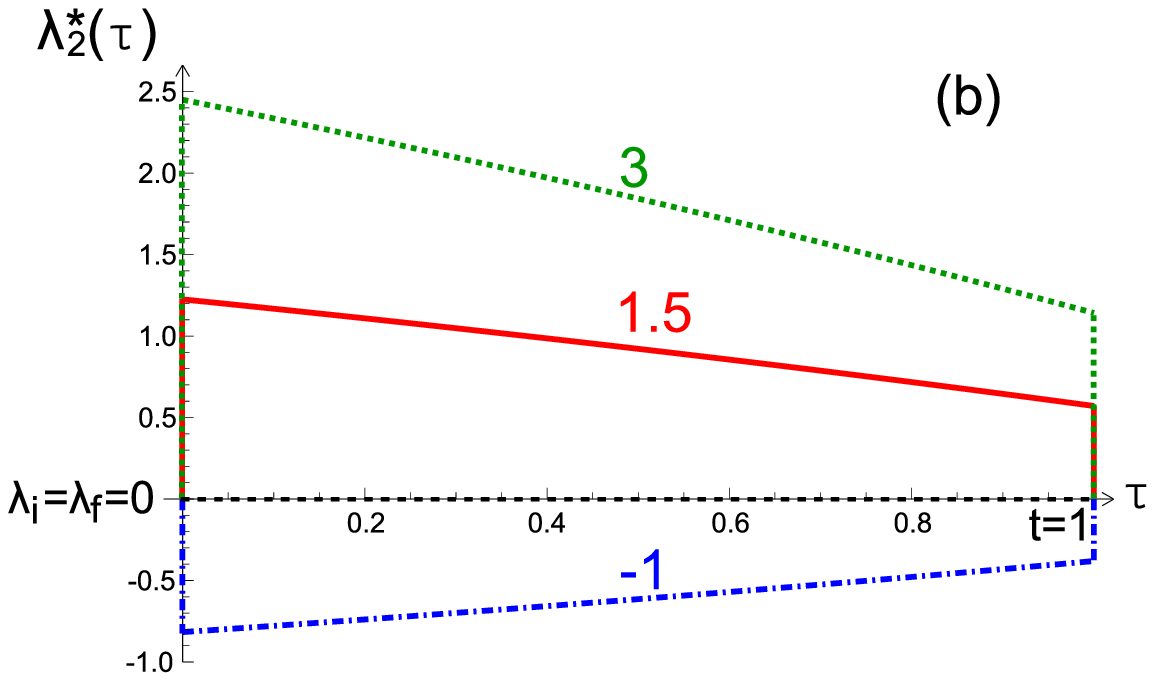}
  \caption{Optimal finite-time protocols for $\kf=1$ and $t=1$. (a) Optimal stiffness $k^*_2(\tau)$ for three different values of $\ym$. (b) Optimal position $\lambda^*_2(\tau)$ of the potential for $\ym = 0.4$, $\lf=0$, and three different measurement outcomes $\xm$. }
  \label{Optprot}
  \end{figure}

If we control not only the position $\lambda(\tau)$ but also the stiffness $k(\tau)$ of the potential, we have to use the full expression \eqref{work} for the mean work. Like we already did to optimize \eqref{workao}, we can use the Euler-Lagrange equations to first optimize the integral terms and show that $b(\tau)$ and $y(\tau)$ must be linear. The total work then reads
  \begin{align}
  W_2 &= \frac{\kf}{2}(b(t)-\lf)^2-\frac{b^2(0)}{2}+ \frac{(b(t)-b(0))^2}{t}\nonumber\\
  &+\frac{\kf}{2}y^2(t)-\frac{y^2(0)}{2}+\frac{(y(t)-y(0))^2}{t}-\logn{\frac{y(t)}{y(0)}}.\label{workinterm}
  \end{align}
Here, $b(0)$ and $y(0)$ are given by eq. \eqref{bzero} and the subscript $_2$ denotes the control of both parameters. We set $\lf=\li=0$ and $\kf=\ki=1$ so that the final potential is equal to the initial one and $\Delta F = 0$. The total work \eqref{workinterm} can be separated in two terms, one depending only on $b(t)$ and the other on $y(t)$. Some arithmetic yields the optimal values for the final distribution,
  \begin{align}
  b^*_2(t) &= \frac{2b(0)}{2+t} \label{bstar2},\\
  y^*_2(t) &= \frac{y(0)+\sqrt{y^2(0)+t(2+t)}}{2+t} \label{ystar2},
  \end{align}
that we can put into eq. \eqref{workinterm} to obtain the optimal work and into eqs. \eqref{fpk} and \eqref{fpl} to get the optimal protocols, shown in fig. \ref{Optprot}.
The protocols show two jumps at $\tau=0$ and $\tau=t$, as previously found in \cite{schm07}. The first jump can be quantified by putting the moments \eqref{bzero} of the initial PDF into eqs. \eqref{fpk} and \eqref{fpl}. This first jump adapts the shape of the potential to the measured position of the particle, depending on the precision $\ym$ of the measurement - the more precise the measurement is ($\ym\to 0$), the sharper the potential after the measurement has to be. The particle is then dragged back closer to the position $\lf=0$ while the potential also flattens towards $\kf=1$. The end jumps vanish in the quasistatic limit $t\to\infty$. 

\begin{figure}[t]
  \centering
  \includegraphics[width=0.8\linewidth]{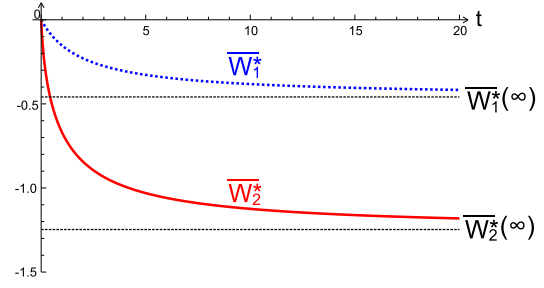}
  \caption{Optimal mean work $\overline{W^*_{1,2}}$ as function of the duration of the process $t$ for $\ym = 0.3$. If the stiffness $k(\tau)$ of the potential is constant, the quasistatic work is given by eq. \eqref{workkonst}. By controlling both $\lambda(\tau)$ and $k(\tau)$, we can extract the work \eqref{workinfo}.}
  \label{Work}
  \end{figure}

If we put the optimal values \eqref{bstar2} and \eqref{ystar2} into eq. \eqref{workinterm} and add the information given by eq. \eqref{kldist}, we obtain
\begin{align}
 W^*_2(\xm)+I(\xm) = &\frac{b^2(0)}{2+t}+\frac{(y_2^*(t)-y(0))^2}{t} \\
  &+\frac{{y_2^*}^2(t)-1}{2}-\ln y_2^*(t) \geq 0 = \Delta F\nonumber
\end{align}
which is always positive since we have $0\leq y_2^*(t)\leq1$. The generalized second law \eqref{2ndlaw} is therefore obeyed, as it was also the case for a constant stiffness. However, equality, $W^*_2(\xm)+I(\xm)=0$, is now reached in the quasistatic limit since $y^*_2(t\to\infty)=1$. We can thus convert all acquired information into work in the quasistatic limit. By averaging $W_2^*(\xm)$ over $\xm$, we obtain the mean total work $\overline{W_2^*}$. Its limit for $t\to\infty$ becomes
  \begin{equation}
  \overline{W_2^*}(t\to\infty) = -\frac{1}{2}\logn{1+\frac{1}{\ym^2}}
  \label{workinfo}
  \end{equation}
and is exactly equal to the opposite of the mutual information \eqref{info} that we acquired through the initial measurement. The mean total work $\overline{W_2^*}$ is always greater than $\overline{W_1^*}$ as shown in Fig. \ref{Work}.

A second interpretation provides further insight into this process. In terms of extracted work, it is equivalent to a process without measurement from an initial potential of stiffness $\ki=1/y^2(0)$ to a potential with $\kf=1$. In this case, the mutual information \eqref{info} is equal to the free energy difference between the final and the initial state which could be extracted through a reversible (quasistatic) driving of the control parameters. Furthermore, the optimal protocol is exactly the same, with the exception of the first jump, if we go without measurement from the initial potential
  \begin{equation}
  V_\text{m}(x,0) \equiv -\logn{\sqrt{2\pi y^2(0)}p(x|\xm)} = \frac{(x-b(0))^2}{2y^2(0)}
  \label{virtual}
  \end{equation}
to the final potential $V_\text{f}(x)=x^2/2$. Thus, the measurement saves us some cost of preparing the system in the initial state \eqref{virtual} since this initial potential is adapted to the measured position $\xm$. It then becomes possible to extract work from the system for any duration $t$ of the protocol.

\section{Cyclic machines}
So far, we have analyzed the optimal protocol following one measurement. In a next step, it is interesting to conceive a periodically working engine which by repeated measurements extracts work. An analytical calculation of the power delivered by such a machine is slightly non-trivial because the PDF for the true (and hence the measured) position of the particle will depend on all previous measurements and protocols. A rough estimate, however, is possible using our results. If only the position of the potential can be controlled, eq. \eqref{workkonst} suggest that one can extract about the mean internal energy per relaxation time of the particle, \textit{i.e.}, a maximum power $P\sim k_BT \mu \ki$ if the units are restored. If in addition the stiffness can also be controlled, the maximum work per measurement according to eq. \eqref{workinfo} increases by a factor given by the logarithm of the precision of the measurement, which leads to the estimate for the maximum power $P\sim k_BT \mu \ki \ln(1/\ym^2)$ for $\ym \ll 1$.

\section{Concluding perspective}
For the paradigmatic case of a colloidal particle bound in a harmonic potential, we have studied how to extract work from a single heat bath using the information gained from a position measurement with finite precision. This work obeys a generalized second law inequality relating extracted work, information and free energies. Equality can only be reached if both center and stiffness of the potential are controlled in a quasistatic process. For finite-time processes the inevitable losses can be minimized by using the calculated optimal protocols. 

Our results should be accessible experimentally by using a colloidal particle in an optical trap, a setup which has previously been used to determine distributions for heat and work that enter the various fluctuation theorems in stochastic thermodynamics. On the theoretical side, it would be interesting to extend our study of the relationship between the concepts of stochastic thermodynamics and those of information theory in finite-time processes beyond the harmonic case to systems with  anharmonic or interacting degrees of freedom.

\bibliographystyle{eplbib}
\bibliography{paper}

\end{document}